\documentstyle[eqsecnum,aps,epsf]{revtex} 

\begin{document}

\twocolumn[\hsize\textwidth\columnwidth\hsize\csname@twocolumnfalse\endcsname

\title{Collapse of attractive Bose-Einstein condensed vortex states in
cylindrical
trap}

\author{Sadhan K. Adhikari}
\address{Instituto de F\'{\i}sica Te\'orica, Universidade Estadual
Paulista, 01.405-900 S\~ao Paulo, S\~ao Paulo, Brazil\\}

\date{\today}
\maketitle
\begin{abstract}

Quantized vortex states of weakly interacting Bose-Einstein condensate of
atoms with attractive interatomic interaction in an axially symmetric
harmonic oscillator trap are investigated using the numerical solution of
the time-dependent Gross-Pitaevskii (GP) equation obtained by  the
semi-implicit
Crank-Nicholson method. Collapse of the condensate is studied in the
presence of deformed traps with a larger frequency along the radial as
well as along the axial directions. The critical number of atoms for
collapse is calculated as a function of vortex quantum $L$. The critical
number increases with angular momentum $L$ of the vortex state but tends
to saturate for large $L$.

{\bf PACS Number(s):  02.70.-c,02.60.Lj,03.75.Fi}

\end{abstract}

\vskip1.5pc]
 \newpage

\section{Introduction}

Recent experiments \cite{1,1a} of Bose-Einstein condensates (BEC) in
dilute bosonic atoms employing magnetic traps at ultra-low temperatures
have intensified theoretical investigations on various aspects of the
condensate \cite{3a,3b,4,11,2}. The properties of the condensate are
usually
described by the nonlinear mean-field Gross-Pitaevskii (GP)  equation
\cite{8}, which properly incorporates the trap potential as well as the
interaction among the atoms.

Two interesting features of BEC are (a) the collapse in the case of 
attractive atomic interaction \cite{1a,2} and (b) the possibility of the
formation of a vortex state in  harmonic traps with  cylindrical
\cite{2a,2b1,2c,2d,2e1} as well
as spherical \cite{2e2} symmetry.

For attractive interatomic interaction \cite{1a,2}, the condensate is
stable for a critical maximum number of atoms.  When the number of
atoms
increases beyond this critical value, due to interatomic attraction the
radius of BEC tends to zero and the maximum  density of the
condensate tends to infinity.
Consequently, the condensate collapses emitting atoms until the number of
atoms is reduced below the critical number and a stable configuration is
reached. With a supply of atoms from an external source the condensate can
grow again and thus a series of collapses can take place, which was
observed experimentally in the BEC of $^7$Li with attractive interaction
\cite{1a}. Theoretical analyses based on the GP equation also confirm
the collapse \cite{2}.

The study of superfluid properties of BEC is of great interest to both
theoreticians \cite{2a,2b1,2c,2d,2e1,2e2,str,w,br}
and experimentalists \cite{exp1,exp2}.  Quantized vortex state
in BEC is intimately connected to the existence of superfluidity. Such
quantized vortices are expected in superfluid He$_{\mbox{II}}$. However,
due to strong interaction between the helium atoms there is no reliable
mean-field description. On the other hand,  a weakly interacting trapped
BEC is well-described
by the mean-field GP equation which  is known to admit vortex solutions
for
a trap with cylindrical symmetry \cite{2a,w}, that  can be studied
numerically. 
This allows for a controlled theoretical study of quantized vortices in
BEC in contrast to  superfluid He$_{\mbox{II}}$.

Many different techniques for
creating
vortex states in BEC have been suggested \cite{2d}, e.g., stirring the
BEC by an external laser at a rate exceeding a critical angular
velocity to create a singly quantized vortex line along the axis 
of rotation \cite{2c}, spontaneous vortex formation
in evaporative cooling \cite{2f}, controlled excitation to an
excited state of atoms \cite{dum}, and rotation of an axially symmetric
trap
\cite{2g}.  
Moreover, quantized vortex states in BEC have been observed
experimentally in coupled BEC comprising of two spin states of $^{87}$Rb
in spherical trap, where angular momentum is generated by a
controlled excitation of the atoms between the two states
\cite{exp2}.  Vortices have also been detected 
in a single-state BEC of $^{87}$Rb  in cylindrical trap, where
angular momentum is generated by a stirring laser beam  
\cite{exp1}. 
After the
possibility of continuously changing the interaction between cold
$^{85}$Rb atoms by
a magnetic-field-induced Feshbach resonance \cite{exp3,hu}, one could
experimentally form  vortex states
in repulsive condensates and study their collapse after transforming
them to
attractive condensates  by such a resonance.
Because of the intrinsic
interest in BEC of vortex states in axially symmetric  traps, in
this work the formation 
of such a BEC is studied using the numerical solution of the
time-dependent GP equation with special
attention to its collapse for attractive interatomic interaction.

In general, a vortex line in a nonrotating trapped BEC is
expected to be nonstationary. However, it is possible to have
dynamically
stable vortex BEC states in a nonrotating trap with low quanta of 
rotational excitation or angular momentum $L$ per particle
\cite{2a,str,w,2g}. Vortex BEC states for large repulsive condensates 
with high  quanta of
rotational 
excitation are expected to  be unstable and decay to vortices with low
quanta  \cite{2c,2d,2e2,br}.
In the absence of vortex,  the stable condensate  in an axially symmetric 
trap  
has a cylindrical shape.  
Such a  BEC has the largest
density on the axis of the  trap. For purely attractive interaction, 
with the increase of the number of atoms the central density of this
condensate 
increases rapidly leading to instability and collapse \cite{2}. 

In the presence of vortex motion the region of largest density of the BEC
with nonzero $L$ 
is pushed away from the central axial region and the atoms have more space
to stabilize. The vortex state of the condensate  in a cylindrical trap 
has the shape of a hollow cylinder with zero density on the axis of
symmetry. Because of larger espacial extension of such a condensate, it can
accommodate a  larger critical number of atoms
before the density increases too high to lead to collapse \cite{2a}.  
  The higher the angular momentum $L$ in a BEC,  the larger is the
critical number of atoms. However,  the increase of this
critical number with
$L$ slows down as $L$ increases.

The present study is performed with the direct numerical solution of the
time-dependent GP equation with an axially symmetric trap. In the
time-evolution of the GP equation the radial and axial variables are dealt
with in two independent steps. In each step the GP equation is solved by
discretization with the   Crank-Nicholson  rule 
complimented by the known boundary
conditions \cite{koo}. We find that this time-dependent
approach leads to good convergence. There are several other
iterative
approaches to the numerical solution of the time-dependent and
time-independent GP equation
for axially symmetric 
\cite{3b,2a,2b1,xx1,2b2} as well as spherically symmetric \cite{3a}
traps. Of
the
time-dependent methods, the approach of Refs. \cite{2b1} uses 
alternative iterations in radial and axial directions as in this study,
whereas Ref. \cite{xx1} does not give the details of the numerical method
employed and  
Ref. \cite{2b2} employs a completely different scheme, e.g., uses
alternative  iterations for the real and imaginary
parts
of the GP equation. However, Refs. \cite{2a,2b1} do not provide enough
details of the numerical scheme. Because of these  a
meaningful comparison of the present method with those of
Refs. \cite{2a,2b1,xx1,2b2} is not possible.

In Sec. II we describe the time-dependent form of
the GP equation including the vortex states for attractive interaction. In
Sec. III we describe the numerical method for solving the time-dependent 
GP equation in some
detail. In Sec. IV we report the numerical results for the collapse of
the BEC with vortex quantum for attractive interaction and finally, in
Sec. V we give a summary of our investigation.

\section{Nonlinear Gross-Pitaevskii Equation}

At zero temperature, the time-dependent Bose-Einstein condensate wave
function $\Psi({\bf r},\tau)$ at position ${\bf r}$ and time $\tau $ may
be described by the self-consistent mean-field nonlinear GP equation
\cite{8}. In the presence of a magnetic trap of cylindrical symmetry this
equation is written as
\begin{eqnarray}\label{a} \biggr[ -\frac{\hbar^2}{2m}\nabla^2
&+& V({\bf r})  
+ gN|\Psi({\bf
r},\tau)|^2\nonumber \\
&-&i\hbar\frac{\partial
}{\partial \tau} \biggr]\Psi({\bf r},\tau)=0.   \end{eqnarray} Here $m$
is
the mass of a single bosonic atom, $N$ the number of atoms in the
condensate, $  V({\bf r}) $ the attractive harmonic-oscillator trap
potential with cylindrical symmetry, 
 $g=4\pi \hbar^2 a/m $ the strength of interatomic interaction, with
$a$ the atomic scattering length. A
positive $a$
corresponds to a repulsive interaction and a negative $a$ to an attractive
interaction. The normalization condition of the wave function is
\begin{equation}\label{4}
 \int d{\bf r} |\Psi({\bf r},t)|^2 = 1.  \end{equation}

The trap potential with cylindrical symmetry may be written as  $  V({\bf
r}) =\frac{1}{2}m \omega ^2(r^2+\lambda^2 z^2)$ where 
 $\omega$ is the angular frequency
of the potential in the radial direction $r$ and 
$\lambda $ is the
ratio of the
axial to radial frequencies. We are using the cylindrical
coordinate system ${\bf r}\equiv (r,\theta,z)$ and in  case of  
cylindrical
symmetry  the wave function is taken to be independent of $\theta$ in the
absence of vortex states of the condensate:
\begin{equation}\label{nang}
\Psi({\bf r},\tau)=\psi(r,z,\tau).
\end{equation}

The GP
equation with a cylindrically symmetric trap  can easily accommodate
quantized vortex
states with rotational motion of the condensate around the $z$ axis 
without any added complication. In such a vortex the atoms flow with
tangential velocity $L\hbar/(mr)$ such that each atom has quantized 
angular momentum
$L\hbar$ along $z$ axis. This corresponds to an angular dependence of
\begin{equation}\label{ang}
\Psi({\bf r},\tau)=\psi(r,z,\tau)\exp (iL\theta)
\end{equation}
 of the wave
function, where $\exp (iL\theta)$ are the circular harmonics in two 
dimensions.   Equation (\ref{nang}) is the zero angular momentum version
of (\ref{ang}). 

Substituting Eq. (\ref{ang}) into 
Eq.  (\ref{a}), one obtains the following GP equation in partial-wave form
with quantized angular momentum $L$ along the $z$ axis 
\begin{eqnarray}\label{c} 
\biggr[
-\frac{\hbar^2}{2m}\biggr(
\frac{1}{r}\frac{\partial }{\partial
r}r\frac{\partial
}{\partial r}&+&\frac{\partial^2}{\partial z^2} -
\frac{L^2}{r^2}\biggr) +
\frac{1}{2}m \omega ^2(r^2+\lambda^2 z^2) \nonumber \\+gN|\psi ({
r,z},\tau)|^2
&-& i\hbar\frac{\partial}{\partial \tau}\biggr] \psi(r,z,\tau)=0,
\end{eqnarray}
with $L=0,1,2,...$ The nonzero values of $L$ corresponds to vortex states. 
The $L^2/r^2$ term in 
Eq. (\ref{c}) is   the vortex contribution to the Hamiltonian of the  GP
equation. This is also the centrifugal barrier term in the
partial-wave linear Schr\"odinger equation. 
 The  limitation to cylindrical symmetry  reduces the GP equation in
three
space dimensions to a two-dimensional partial differential equation. We
shall study
numerically this equation in this paper to understand the effect of the 
 $L^2/r^2$ term to collapse in the case of attractive atomic interaction. 

It is convenient to use dimensionless variables
defined by $x =\sqrt 2 r/l$,  $y=\sqrt 2 z/l$,   $t=\tau \omega, $ and  
\begin{equation}\label{wf}
\phi(x,y,t)\equiv 
\frac{ \varphi(x,y,t)}{x} =  \sqrt{\frac{l^3}{2\sqrt 2}}\psi(r,z,\tau),
\end{equation} 
where
$l\equiv \sqrt {\hbar/(m\omega)}$. Although $\phi(x,y,t)$ is the
dimensionless wave function,  for calculational purpose we shall
be using $\varphi(x,y,t)$ in the following. 
 In terms of these
 variables Eq. (\ref{c}) becomes
\begin{eqnarray}\label{d}
\biggr[ -\frac{\partial^2}{\partial
x^2}&+&\frac{1}{x}\frac{\partial}{\partial x} -\frac{\partial^2}{\partial
y^2}
+\frac{L^2}{x^2}
+\frac{1}{4}\left(x^2+\lambda^2 y^2-\frac{4}{x^2}\right) \nonumber \\
&+& 8 \sqrt 2 \pi   n\left|\frac {\varphi({x,y},t)}{x}\right|^2 
-i\frac{\partial
}{\partial t} \biggr]\varphi({ x,y},t)=0, 
\end{eqnarray}
where
$ n =   N a /l.$ 
A reduced number of particles is defined as
$|n|$.
The normalization condition (\ref{4}) of the wave
function become \begin{equation}\label{5} {2\pi} \int_0 ^\infty
dx \int _{-\infty}^\infty dy|\varphi(x,y,t)|
^2 x^{-1}=1.  \end{equation}
However, physically it could be more  interesting  to define   the
 reduced number of particles in terms of a geometrically averaged
frequency $
\omega_0 = \lambda^{1/3} \omega$ and a length $l_0=\sqrt{\hbar /m
\omega_0}$, so that a new reduced number $k(\lambda )$ 
is defined via \cite {xx1} 
\begin{equation} \label{kk}
k(\lambda) \equiv \frac {N|a|}{l_0} = n \lambda ^{1/6}.
\end{equation}
We shall study this number in the present paper.

For a stationary solution the time dependence of the wave function is
given by $\varphi(x,y,t) = \exp(-i\mu t ) \varphi(x,y)$
where $\mu$ is the chemical potential of the condensate in units of
$\hbar\omega$. If we use this
form of the wave function in Eq. (\ref{d}), we obtain the following
stationary nonlinear time-independent GP equation \cite{8}: 
\begin{eqnarray}\label{dx}
\biggr[ -\frac{\partial^2}{\partial
x^2}&+&\frac{1}{x}\frac{\partial}{\partial x} -\frac{\partial^2}{\partial
y^2}+\frac{L^2}{x^2}
+\frac{1}{4}\left(x^2+\lambda^2 y^2-\frac{4}{x^2}\right) \nonumber \\
&+&8\sqrt 2 \pi n\left|\frac {\varphi({x,y})}{x}\right|^2 
-\mu  \biggr]\varphi({ x,y})=0. 
\end{eqnarray}
Equation (\ref{dx}) is the stationary version of the
time-dependent Eq. (\ref{d}).
However,  Eq.  (\ref{d}) is equally
useful for obtaining a stationary solution with trivial time dependence as
well as for studying evolution processes with explicit time dependence and we shall be
directly solving Eq. (\ref{d}) numerically in this paper.

Two  interesting properties of the condensate wave function are the 
mean-square
sizes in the radial and axial directions  defined, respectively,  by 
\begin{equation}\label{7a}  \langle x^2
\rangle= 2\pi \int_0
^\infty dx \int _{-\infty}^\infty dy x |\varphi(x,y,t)| ^2,
\end{equation}
and 
\begin{equation}\label{7b} \langle y^2
\rangle= 2\pi
\int_0
^\infty dx \int _{-\infty}^\infty dy x^{-1}y^2 |\varphi(x,y,t)| ^2.
\end{equation}

\section{Numerical Method}

To solve the time-independent GP equation we need the boundary conditions
of the wave function as $x \to 0$ and $\infty$ and $|y| \to \infty$. For a
confined condensate, for a sufficiently large $x$ and $|y|$,
$\varphi(x,y)$ must vanish asymptotically. Hence the cubic nonlinear term
can eventually be neglected in the GP
equation for large $x$ and $|y|$ and Eq. (\ref{dx}) becomes
\begin{eqnarray} \label{dy} \biggr[ -\frac{\partial^2}{\partial
x^2}&+&\frac{1}{x}\frac{\partial}{\partial x} -\frac{\partial^2}{\partial
y^2}+\frac{L^2}{x^2} +\frac{1}{4}\left(x^2+\lambda^2
y^2-\frac{4}{x^2}\right)  \nonumber \\ &-& \mu 
\biggr]\varphi({ x,y})=0.  
\end{eqnarray} This is the equation for the free oscillator with
cylindrical symmetry in partial-wave form.  The wave
function for a general state 
of this oscillator  and the corresponding
energy are given, respectively, by \cite{pw}
\begin{equation} \varphi({
x,y})= {\cal N} x e^{-(x^2+\lambda y^2)/4}F_{|L|,n_x}\left( \frac{x}{\sqrt
2}\right)H_{n_y}\left(\frac{y\sqrt
\lambda}{\sqrt 2}\right),
\end{equation} 
and 
\begin{equation} 
\mu = \left(1+|L|+n_x\right)+\left(n_y+\frac{1}{2}\right)\lambda,
\end{equation} 
with $L= 0, \pm 1, \pm 2, ...$, $n_x= 0,2,4,...$, and $n_y=0,1,2,...$
Here $H_{n_y}$ is the usual Hermite polynomial, and $F_{|L|,n_x}$
is another polynomial defined recursively \cite{pw,hp}, 
and ${\cal N}$ is
the 
normalization.  The first few of these polynomials are: 
$H_0(\xi)=1, H_1(\xi)=2\xi, H_2(\xi)=(4\xi^2 -2), H_3(\xi)=\xi(8\xi^2
-12),
F_{0,0}(\xi)=1,
F_{1,0}(\xi)= \xi,   
F_{2,0}(\xi)= \xi^2, F_{0,2}(\xi)= (1-\xi^2), F_{3,0}(\xi)= \xi^3,
F_{1,2}(\xi)= \xi(\xi^2-4)$
etc. \cite{hp}.
In this paper we shall be interested in angular momentum
(vortex) excitation, opposed to radial excitation via $n_x$ or
axial excitation via $n_y$,
 of the following normalized ground
state wave
function for  $n_x=n_y=0$    
\begin{equation}\label{ho} \varphi({
x,y})= \left( \frac{ \lambda
}{2^{2L+3}\pi^3(|L|!)^2}\right)^{1/4}x^{1+|L|} e^{-(x^2+\lambda y^2)/4},
\end{equation} 
with energy 
\begin{equation} 
\mu = 1+|L|+\frac{1}{2}\lambda.
\end{equation} 
Solution (\ref{ho}) of Eq. (\ref{dy}) is a good starting
point for
a iterative method for solving the time-dependent GP equation
(\ref{d}) for small values of
nonlinearity $n$ as in this paper. Alternatively, to solve the GP equation
for large nonlinearity $n$, one may start with the Thomas-Fermi
approximation for the wave function obtained by setting all the
derivatives in the GP equation to zero \cite{11}, which is a good
approximation for large nonlinearity.

Next we consider Eq. (\ref{d}) as $x\to 0$. The nonlinear term approaches
a constant in this limit because of the regularity of the wave function at
$x=0$. Then one has the following condition \begin{equation}
\varphi(0,y)=0, \label{11} \end{equation} as in the case of the harmonic
oscillator wave function (\ref{ho}).  Both the small- and large-$x$
behaviors of the wave function are necessary for a numerical solution of
the time-dependent GP equation (\ref{d}). The large-$x$ and large-$|y|$
behaviors of the wave function are given by Eq. (\ref{ho}), e.g.,
\begin{eqnarray}\label{11x} \lim_{x\to \infty} \varphi(x,y) \to e^
{-x^2/4}, \\ \lim_{|y|\to \infty} \varphi(x,y) \to  e^ {-
\lambda y^2/4}. \label{11y} \end{eqnarray}

A convenient way to solve Eq. (\ref{d}) numerically is to discretize it in
both space and time and reduce it to a set of algebraic equations which
could then be solved by using the known asymptotic boundary conditions.
The method of solution using one space derivative is well under control 
\cite{3a,koo}.
The GP equation  (\ref{d}) can be  written formally as
\begin{equation}
i\frac{\partial}{\partial t} \varphi =H \varphi,
\end{equation}  
where $H$ is the time-independent quantity in the square brackets of
Eq. (\ref{d}). The integration in time is effected via the
following semi-implicit Crank-Nicholson algorithm \cite{koo}
\begin{equation}
\frac{ \varphi^{n+1}- \varphi^n}{-i\Delta } = \frac{1}{2}H(
\varphi^{n+1}+ \varphi^n),\label{a1}
\end{equation}
where $\Delta $ is the constant time step used to calculate the time
derivative,  $ \varphi^n$ is the discretized wave function at time
$t_n=n\Delta $, and where the space variables $x$ and $y$ are suppressed.
The derivatives in the operator $H$ are discretized by the finite
difference scheme \cite{koo}.
The formal solution to Eq. (\ref{a1}) is given by
\begin{equation}\label{a2}
 \varphi^{n+1}=\frac{1-i\Delta H/2          }{ 1+i\Delta H/2 }
\varphi^{n}
\end{equation}
so that if $  \varphi^{n}   $ is known at time $t_n$ one can find
$\varphi^{n+1}$ at the next time step  $t_{n+1}$. This procedure 
is used
to solve the GP equation involving one space variable \cite{3a}. In that
case after
proper discretization in space using a finite difference scheme
Eq. (\ref{a2})    becomes a tridiagonal set of equations in discrete space
observables at time $t_{n+1}$ which is solved by Gaussian elimination
method and back substitution \cite{koo} using the known boundary
conditions (\ref{11}), (\ref{11x}), and (\ref{11y}). Unfortunately, a similar
straightforward discretization of Eq. (\ref{d}) in
two space observables using a finite difference scheme in this case does
not lead to a tridiagonal set of equation but rather to a unmanageable
set
of equations \cite{koo}. 

To circumvent this problem 
the full $H$ operator in this case is conveniently broken up
into radial and axial  components $H_x$ and $H_y$, respectively,  where
$H_x$ contains the terms
dependent on $x$ and  $H_y$ the terms dependent on $y$ with the nonlinear
term $8\sqrt 2 \pi n|\varphi(x,y)/x|^2$ involving both $x$ and $y$
contributing
equally to
both. Specifically, we take
\begin{equation}
H_x=-\frac{\partial^2}{\partial x^2}+\frac{1}{x}\frac{\partial}{\partial x}
+\frac{L^2-1}{x^2}+\frac{x^2}{4}+4\sqrt 2 \pi  n \left|\frac{\varphi(x,y,t)
}{x}
  \right|^2
\end{equation}
\begin{equation}
H_y=-\frac{\partial^2}{\partial y^2}+\frac{\lambda^2 y^2}{4}+4\sqrt 2 \pi
n
\left|\frac{\varphi(x,y,t)
}{x}
  \right|^2,
\end{equation}
with $H= H_x+H_y$.
However, the numerical result of the present scheme is independent of a specific
breakup. 

The procedure is then to define the unknown wave function on a
two-dimensional mesh in the $x-y$ plane. The time evolution is then
performed in two steps. 
First the time evolution is
effected using the operator $H_x$  setting $H_y=0$ along lines of constant
$y$ with $i\partial \varphi /\partial t = H_x \varphi$.  Next the  time evolution is
effected using the operator $H_y$ setting $H_x=0$ along lines of constant
$x$  with $i\partial \varphi /\partial t = H_y \varphi$. This procedure
is repeated alternatively. This scheme is conveniently
represented in terms of an  auxiliary function
$\varphi^{n+\frac{1}{2}}$ by
\begin{equation}
 \varphi^{n+1}=\frac{1-i\Delta H_y/2          }{ 1+i\Delta H_y/2 }
\varphi^{n+\frac{1}{2}},
\quad 
 \varphi^{n+\frac{1}{2}}=\frac{1-i\Delta H_x/2          }{ 1+i\Delta
H_x/2 }
\varphi^{n}, 
\end{equation}
so that 
\begin{equation}
 \varphi^{n+1}=\frac{(1-i\Delta H_y/2)          }{ (1+i\Delta H_y/2 )}
\frac{(1-i\Delta H_x/2 )         }{ (1+i\Delta
H_x/2) }
\varphi^{n},\label{a3}
\end{equation}
where $n=0,1,2,...$ denotes the number of iterations.
For a small time step $\Delta$, if we neglect terms quadratic in $\Delta$,
Eq. 
(\ref{a3})
is equivalent to (\ref{a2}). Hence for numerical purpose we have been able
to reduce the
 GP equation in two space dimensions, $x$ and $y$, 
into a series of GP equations in one space variable, either $x$ or $y$.
The GP equations in one space variable can be dealt with numerically   
 in a standard fashion using  Crank-Nicholson discretization and
subsequent solution by the Gaussian elimination method. This scheme is
stable  independent of time step employed.

The time-dependent GP equation (\ref{d}) is solved by time iteration by
mapping the solution on a two-dimensional grid of points $N_x\times N_y$
in $x$ and $y$.
First Eq. (\ref{d}) with $H_x$ is discretized using the following finite
difference
scheme along the $x$ direction within the semi-implicit Crank-Nicholson
rule
\cite{yyy}:
\begin{eqnarray}\label{kn1}
\frac{i(\varphi_{j,p}^{n+1}-
\varphi_{j,p}^n)}{\Delta}=-\frac{1}{2h^2}\biggr[(\varphi^{n+1}_{j+1,p}-2\varphi_{j,p}
^{n+1}
+\varphi^{n+1}_{j-1,p})\nonumber \\
+(\varphi^{n}_{j+1,p}-2\varphi_{j,p}^{n}
+\varphi^{n}_{j-1,p})\biggr]\nonumber \\ 
+
\frac{1}{4x_jh}\left[(\varphi^{n+1}_{j+1,p}-\varphi^{n+1}_{j-1,p}) 
+(\varphi^{n}_{j+1,p}-\varphi^{n}_{j-1,p}) 
\right]     \nonumber \\
+\biggr[\frac{x_j^2}{8}+\frac{L^2-1}{2x_j^2}+2\sqrt 2\pi n
\frac{|\varphi^n_{j,p}|^2}{x_j^2}\biggr]
(\varphi_{j,p}^{n+1}+\varphi_{j,p}
^n),                                 
\end{eqnarray}
where the discretized wave function 
$\varphi^n_{j,p}\equiv \varphi(x_j,y_p,t_n)$ refers to a fixed  $y=y_p=ph,
p=1,2,...,N_y$ at different 
$x=x_j=jh,
j=1,2,...,N_x,$ and $h$ is the space step.
This
procedure results in a series of tridiagonal sets of equations 
(\ref{kn1}) in $\varphi^{n+1}_{j+1,p}$,  $\varphi^{n+1}_{j,p}$, and
$\varphi^{n+1}_{j-1,p}$ at time $t_{n+1}$ for
each  $y_p$, which are solved by Gaussian elimination and back
substitution \cite{koo} starting with the initial harmonic oscillator
solution (\ref{ho}) at $t_0=0$ and $n=0$.  
Then Eq. (\ref{d}) with $H_y$ is
discretized using the following finite difference scheme along the $y$
direction:
\begin{eqnarray}\label{kn2}
\frac{i(\varphi_{j,p}^{n+1}-
\varphi_{j,p}^n)}{\Delta}=-\frac{1}{2h^2}\biggr[(\varphi^{n+1}_{j,p+1}-2\varphi_{j,p}
^{n+1}
+\varphi^{n+1}_{j,p-1})\nonumber \\
+(\varphi^{n}_{j,p+1}-2\varphi_{j,p}^{n}
+\varphi^{n}_{j,p-1})\biggr]\nonumber \\ 
+\left[\frac{\lambda^2y_p^2}{8}+2\sqrt 2\pi n
\frac{|\varphi^n_{j,p}|^2}{x_j^2}
\right] (\varphi_{j,p}^{n+1}+\varphi_{j,p} ^n),                            
\end{eqnarray}
where now $\varphi^n_{j,p}$ refers to a fixed $x_j=jh$ for all 
$ y_p = ph.$ 
Using the solution obtained after $x$ iteration as input, the discretized
tridiagonal equations (\ref{kn2}) along the $y$ direction for constant $x$
are solved
similarly. This two-step
procedure corresponds to a full iteration of the GP equation and the
resultant solution corresponds to time $t_1=\Delta$, and $n=1$.  This
scheme is repeated for about 500  times to yield the final solution
of
the GP equation. The normalization condition (\ref{5}) is preserved 
during time iteration due to the unitarity of the time-evolution operator. 
However, it is convenient to reenforce it numerically after each iteration
in order to maintain a high level of precision. Also, the solution at each
time step will satisfy the boundary
conditions (\ref{11}), (\ref{11x}), and (\ref{11y}).  At each 
iteration the strength of the non-linear term is increased by a small
amount so that after about 500  time iterations the full strength is
attained and the required solution of the GP equation obtained. The
solution so obtained is iterated several times (between 20 to 50
times) until an
equilibrated final result is obtained. This solution is the ground
state of the condensate corresponding to the specific nonlinear constant
$k$ and $L$.  

We found the convergence of the two-step iteration scheme to be fast for
small $|n|$.  
However, the final convergence of the scheme breaks down if $|n|$ is too
large.  For an attractive interaction there is no such problem as the GP
equation does not sustain a large nonlinearity $|n|$.
  Typical values of the parameters used in this paper for 
discretization along $x$ and $y$ directions are $N_x=400$,
$N_y=800$, respectively, with  $x_{\mbox{max}} = 8$, $|y|_{\mbox{max}} =
8$, and $\Delta
=0.05$ for $\lambda >0.5$. For smaller $\lambda  (<0.5)$  the wave
function extends
further  along the $y$ axis and  larger  $|y_{\mbox{max}}| $ and
$N_y=800$
are to be
employed for obtaining converged result.  
The above choice of parameters  corresponds to a typical
space step of $h=0.02$ along both radial
and axial directions.   
These parameters were obtained after
some
experimentation and  are found to lead to good convergence.

As the time dependence of the stationary states is trivial $-$
$\varphi(x,y,t)= \varphi(x,y) \exp(-i\mu t)$ $-$ the chemical potential
$\mu$ can be obtained from the propagation of the converged ground-state
solution at two times, e.g., $\varphi(x,y,t_n)$ and
$\varphi(x,y,t_{n+n'})$.  From the numerically obtained ratio
$\varphi(x,y,t_n)/\varphi(x,y,t_{n+n'})= \exp(i\mu n'\Delta)$, $\mu$ can
be
obtained as the time step $\Delta $ is known. In the calculation of $\mu$
an average over relatively large values of $n'$ leads to stable result.

\section{Numerical Result}

Using the numerical method described in Sec. III we present results for
the numerical solution of the time-dependent GP equation in the following
for attractive interatomic interaction with special attention to the
collapse of the condensate. To assure that we are on the correct track,
using the present program first we solved the GP equation for the
spherically symmetric case with $\lambda = 1$ and $L=0$, and compare with the
calculation of Ref. \cite{xxx}. As an additional check we also solved the
GP equation in two space dimensions with $\lambda =0$ and without the
$d^2/ dy^2$ term in Eq. (\ref{d}) and compare with the calculation of Ref.
\cite{yyy}.  In both cases the present calculation agrees with these
previous ones.

Before describing the results for nonzero $L$ first we compare the present
results for $L=0$ with those of Ref.  \cite{xx1} for a cylindrically
symmetric trap. For the spherically symmetric case $\lambda =1$, and the
critical number $k_c(\lambda)$ of Eq. (\ref{kk}) for collapse is found to be
0.575 in agreement with Refs. \cite{11,xx1,xxx}. In a recent
experiment using $\lambda = 0.3919$, the critical reduced number for
collapse for an attractive condensate of $^{85}$Rb atoms formed using a
Feshbach resonance
was found  to be $k_c=0.459
\pm 0.012\pm 0.054$ \cite{yy1}. In their
calculation Gammal et al \cite{xx1} obtained $k_c=0.550$ for $\lambda =
0.3919$. In the present calculation we obtain $k_c=$
0.553 in excellent agreement with Ref. \cite{xx1} using an entirely
different
numerical routine.  However, the disagreement with the experimental result 
\cite{yy1} remains.  We also calculated the critical number $k_c(\lambda)$ for
some other values of $\lambda$. For $\lambda =5, 2, 0.2$ we
obtain $k_c= 0.50, 0.56$, and 0.52,   respectively, compared to 0.498,
0.561,
and 0.509 obtained in  Ref.  \cite{xx1}.  The small difference between the
results of the two
calculations
 seems to be a consequence of numerical error.  Also, as in Ref.  
\cite{xx1} we note that for $\lambda$ not so different from unity
($5>\lambda>0.2$) the critical reduced number for collapse $k_c(\lambda)$
satisfies $k_c(\lambda)\approx k_c(1/\lambda)$, and attains a maximum at
$\lambda =1$ corresponding to the spherically symmetric situation.
However, this symmetry is broken for large values of $\lambda $,
e.g., for $\lambda >5 $ while 
 we have $k_c(\lambda)< k_c(1/\lambda)$.  Moreover, we
find in the following that this symmetry is also broken for nonzero $L$,
where, however, for $\lambda >1$   $k_c(\lambda)>
k_c(1/\lambda)$.

Next we comment on the discrepancy between the experimental
critical number of atoms for collapse for an  attractive BEC of
$^{85}$Rb atoms formed using a Feshbach resonance \cite{yy1} on one hand
and the theoretical results of Ref.  \cite{xx1} and the present
calculation on the other hand. In view of the success of the mean-field GP
equation to explain many stationary results and time-evolution phenomena
of the attractive BEC of $^7$Li atoms with an almost spherical trap
\cite{1a,2}, it seems that this description is perfectly appropriate for
attractive condensates. Hence, we do not believe that a relatively small
deviation from spherical symmetry as in the experiment of Ref. \cite{yy1}
would invalidate the applicability of the GP equation to an attractive
condensate.  Whether the inclusion of higher order interaction terms in
the mean-field GP equation could account for the observed data \cite{xx1}
yet remains to be established. 
To resolve the discrepancy we advocate further experimental
study of collapse for attractive condensates after changing the trap
symmetry ($\lambda$).

After the above preliminary comparative study, we present results for the
numerical solution of 
the GP equation (\ref{d}) for nonzero 
$L=0,1,2,..,8$ and   $\lambda =\sqrt 8$ and ${1/\sqrt 8}$  for different
 $k(\lambda)$. We recall that  $\lambda =\sqrt 8$ corresponds to 
the experiment of Ensher et al. \cite{1} for the BEC of $^{87}$Rb atoms.  
These two
possibilities of $\lambda$ correspond to axial compression ($\lambda > 1$) 
and elongation  ($\lambda < 1$) 
of the condensate, respectively.  For each $L$ we increase $k$ from
$0$ and calculate the chemical potential $\mu$. With the increase of
$k$ the wave function becomes more and more localized in space and beyond
a certain value of $k$ the density at the peak of the wave-function
diverges 
 and no stable normalizable solution of the GP equation
with a well defined $\mu$ can be obtained.

In Fig. 1 we plot $\mu$ vs.  $k(\lambda)$ for $\lambda = \sqrt 8$ and
$1/\sqrt 8$ for different $L$. We
also exhibit the result for the spherically symmetric case $\lambda = 1$
($L=0$)
for comparison.  The curves are plotted for all allowed values of $k$ for the
ground state in each case. The curves go up to a maximum critical value $k_c$ 
of
$k$ which defines the  critical number $N_c$ of atoms in that particular
case via $k_{c}= N_c |a|/l_0$. We find that (i) $k_c$ for a particular
$\lambda$ increases with $L$ and that (ii)  $k_c$ for a particular nonzero $L$
increases as $\lambda $ increases from $ {1/\sqrt 8}$ to $\sqrt 8$, which
demonstrates the
breakdown of the numerically noted symmetry $k_c(\lambda)\approx
k_c(1/\lambda)$
for $L=0$. To demonstrate these two effects in an explicit fashion we plot in
Fig. 2 $k_c$ vs. $L$ for $\lambda = \sqrt 8 $, 1, and $ {1/\sqrt 8}$. The
three 
curves intersect
approximately 
at $L=0$ which demonstrates that $k_c(\lambda= \sqrt 8)\approx k_c
(\lambda= {1/\sqrt 8})< k_c(\lambda= 1)$
for
$L=0$, with  $k_c(\lambda= \sqrt 8)=0.54$,  $k_c(\lambda=1/ \sqrt
8)=0.55$, and  $k_c(\lambda= 1)=0.575$. 
However,  this symmetry is broken for non-zero $L$ while  $k_c(\lambda=
\sqrt 8)
>k_c(\lambda= 1)> k_c
(\lambda= {1/\sqrt 8})$. The critical number $k_c(\lambda)$ increases with $L$
for all $\lambda$, and we see from Fig. 2 that this rate of increase slows down
as $L$ increases. 


 
 

In Figs. 3 and 4 we plot the wave function
$|\phi(x,y)|\equiv |\varphi(x,y)/x|$ in dimensionless
variables of Eq. (\ref{wf}).  In Figs. 3 (a) $-$ (c) we show the wave
function 
for $\lambda = {1/\sqrt 8}$ and $L=0,2$ and 4, respectively, where the
parameter $k$ is chosen to be very close to the critical value $k_c$ for
collapse.  The nature of the wave function is qualitatively different for
zero and nonzero  $L$. For $L=0$ the wave function is peaked on
the $y$ axis;  whereas for nonzero $L$ it is zero on the $y$ axis and is
peaked at some finite $x$.  In all cases the peak is sharp and the density
of atoms is very large on the peak. The BEC collapses with a slight
increase in the parameter $k$.  For smaller $k$ the wave function has a
much broader maximum. When $k$ approaches $k_c$ a sharp maximum of the
wave function appears very rapidly. To illustrate this in Fig. 3 (d) we
plot the $L=2$ wave function for $k=2.5$. If we compare this with the wave
function of Fig. 3 (b) for $L=2$ and $k=2.58\approx k_c$, the change in
the shape is explicit.

In Figs. 4 (a) $-$ (e) we plot the wave function for $\lambda = {\sqrt 8}$
and for $L=0,2,4,6$, and 8, respectively, for $k\approx k_c$.  If we
compare
Figs. 3 with 4 for same $L$ we find that for $\lambda = {1/\sqrt 8}$ the wave
functions extend over a larger region along the $y$ axis compared to those for
$\lambda =\sqrt 8$. This is apparent if we compare Fig. 3 (a) with 4 (a), and
is expected as $\lambda = \sqrt 8$ corresponds to a stronger harmonic
oscillator potential in the $y$ direction responsible for axial
compression.
From Figs. 3 and 4 we find that for both $\lambda$, the peak in the wave
function moves further away from the $y$ axis as $L$ increases.

To understand some aspects of the  variation of $k_c$ with $L$ and
$\lambda$ exhibited in 
Fig. 2, we plot in
Figs. 5 (a) and (b) the mean-square sizes $\langle x^2\rangle$ and
$\langle
y^2\rangle$ vs. $k$ for different $L$ and for $\lambda =1/\sqrt 8$ and
$ \sqrt 8$,
respectively. The results for vortex states ($L>0$) in the spherically
symmetric case with $\lambda =1$, 
remain between  the those for $\lambda =1/\sqrt 8$ and $\sqrt 8$ and are not
explicitly shown here.
For nonzero $L$ the system acquires a positive
rotational energy $L^2/x^2$ which allows it to move away from the
axial direction $y$. For $L=0$ the region of highest density is the $y$ axis.
For $L\ne 0$ the density  is zero on the $y$ axis and has a
maximum 
at some finite $x$. Consequently, the condensate has the shape of a hollow
cylinder.
Because of vortex motion the condensate swells and has
more space to stabilize. Hence for $L>0$ the density does not go to an
unbearable level with the same number of atoms as for $L=0$, and $k_c$
increases with $L$ for all  $\lambda $. However, for all
$L$ and $\lambda$ with the increase of nonlinearity $k$ (or $n$)  in the GP
equation (\ref{d}), the attractive nonlinear interaction term takes control
and eventually the mean-square sizes ($\langle x^2 \rangle$ and $\langle
y^2\rangle $) reduce as can be seen from Figs. 5. This eventual
shrinking
in size with the
increase of the number of atoms for all $L$ and $\lambda$
together with the outward push due to vortex
motion for nonzero $L$ takes the density of the BEC at the maximum of
the wave function to an unbearably high
level at some critical value $k_c$ of $k$ leading to collapse.

Although, for a fixed $\lambda$,   $k_c$ increases with 
$L$, 
the rate of increase slows down for large $L$. As  $k$ (or $n$)
increases
sufficiently for large $L$ ($>8$),  the nonlinear 
term containing $n$ becomes the
deciding
factor in the GP equation and the $L^2/r^2$ term starts to play a
secondary role. 
Consequently, the increase in the critical number $k_c$  with $L$ slows down as
$L$
increases and the number $k_c$  tends to saturate  as
can be seen clearly in Fig. 2. In all  cases ($\lambda = \sqrt 8$, 1 and
$  {1/ \sqrt 8}$) this
tendency of 
saturation is visible beyond $L =4$.

\section{Summary}

In this paper we present a numerical study of the time-dependent
Gross-Pitaevskii equation under the action of a harmonic oscillator trap
with cylindrical symmetry with attractive interparticle interaction to
obtain insight into the collapse of vortex states of BEC.  The
time-dependent GP equation is solved iteratively by discretization using a
two-step Crank-Nicholson scheme. We obtain the boundary conditions
(\ref{11}), (\ref{11x}), and (\ref{11y})  of the solution of the
dimensionless GP equation (\ref{d})  and use them for its solution. The
solution procedure is applicable for both attractive and repulsive atomic
interactions as well as for both stationary and time evolution
problems. It
is expected that numerical difficulty should appear for large nonlinearity
or large values of reduced number of particles $k$ and large vortex
quantum $L$.  For medium nonlinearity, as in this paper, the accuracy of
the time-independent method can be increased by reducing the space step
used in discretization.

The ground-state wave function for each $L$ is found to be sharply
peaked
for attractive interatomic interaction with the parameters set close 
to those  for collapse. 
In the case of an attractive interaction, the mean square sizes $\langle
x^2
\rangle$ and 
$\langle y^2  \rangle$ 
decrease as  the number of particles in the condensate increases towards the
critical number for collapse. Consequently, the density increases rapidly
signaling the on-set of collapse beyond a critical reduced number $k_c$.

The presence of the quantized vortex states increases the
stability of the BEC with attractive interaction. The critical number 
 $k_c(\lambda)$ for $L=0$ is largest in the spherically symmetric
case 
$\lambda =1$. For vortex states ($L\ne 0$),  $k_c(\lambda)$ increases with
$\lambda$.  
As the 
vortex quantum $L$ increases, $k_c$ also
increases. However, in the present calculation a tendency of saturation in the
value of 
$k_c$  is noted with the increase of
$L$. As the parameter $n$ or $k$ in the GP
equation  increases, the nonlinear term starts to play the dominating role
in the GP equation compared to the angular momentum term
$L^2/x^2$. Once this happens, the rate of increase  of $k_c$  with $L$
slows down
and it is not unlikely that  the critical number 
attains a limiting maximum value  for a larger $L (> 8)$ than those
considered in
this paper. This and other investigations on the collapse of vortex states are
welcome in the future.

The work is supported in part by the Conselho Nacional de Desenvolvimento
Cient\'\i fico e Tecnol\'ogico and Funda\c c\~ao de Amparo \`a Pesquisa do
Estado de S\~ao Paulo of Brazil.

\vskip 1cm

{\bf Figure Caption:}

1. Chemical potential $\mu$ vs. reduced number $k$ for different $\lambda$ and
$L$. The curves are labeled by their respective $L$ value. 

2. Critical reduced number $k_c$ vs. $L$ for $\lambda =\sqrt 8$ (full line
with
$\times$), 1 (dashed-dotted line with ${\bf +}$)  and $1/\sqrt 8$ (dashed
line
with $\star$). The lines are
polynomial fits to
the points.

3. The wave function $|\phi(x,y)|\equiv |\varphi(x,y)/x|$ vs. $x$ and $y$
for $\lambda ={1/\sqrt 
8}$
and for 
(a) $L=0$, $k= 0.54$
(b) $L=2$, $k= 2.58$   
(c) $L=4$,  $k= 4.00$    and 
(d) $L=2$,  $k= 2.50$.

4. Same as figure 3  for $\lambda =\sqrt 8$
and for 
(a) $L=0$, $k= 0.53$,
(b) $L=2$, $k= 3.15$,
(c) $L=4$,  $k= 5.13$,  
(d) $L=6$, $k= 6.71$, and 
(e) $L=8$, $k= 8.12.$

5. Mean square sizes $\langle x^2 \rangle$ (full line)  and  $\langle y^2
\rangle$  (dashed line)
vs. reduced number $k$ for (a) $\lambda =1/\sqrt 8$  and 
(b) $\lambda = {\sqrt 8}$.


\begin{references} 
\bibitem{1}M. H. Anderson, J. R. Ensher, M. R. Matthews,
C. E. Wieman, and E. A. Cornell, Science {\bf 269}, 198 (1995);  J. R.
Ensher, D. S. Jin, M. R. Matthews, C. E. Wieman, and E. A.  Cornell, Phys.
Rev. Lett.  {\bf 77}, 4984 (1996);  K. B. Dadic, M. O. Mewes, M. R.
Andrews, N. J. van Druten, D. S. Durfee, D. M. Kurn, and W. Ketterle,
{\it ibid.}  {\bf 75}, 3969 (1995);  D. G. Fried, T. C. Killian, L.
Willmann, D. Landhuis, S. C.  Moss, D. Kleppner, T. J. Greytak, {\it
ibid.}   {\bf 81}, 3811 (1998); 
F. Pereira Dos Santos, J. L\'eonard, Junmin Wang, C. J. Barrelet,
F. Perales, E. Rasel, C. S. Unnikrishnan, M. Leduc, and C.
   Cohen-Tannoudji, {\it ibid.}   {\bf 86}, 3459 (2001). 

 
\bibitem{1a} C. C. Bradley, C. A. Sackett,
J. J.  Tollett, and  R. G. Hulet, 
{Phys. Rev. Lett.}  {\bf 75}, 1687 (1995);
C. A.  Sackett, H. T. C.   Stoof,  and  R. G. Hulet, 
{\it ibid.}    {\bf 80},  2031 (1998); 
C. C. Bradley, C. A. Sackett, and R. G. Hulet, {\it ibid.}  
{\bf 78}, 985 (1997);
J. M. Gerton, D. Strekalov,
I. Prodan, and R. G. Hulet, Nature (London) {\bf 408},
692 (2000).



\bibitem{3a}M. Edwards and K. Burnett, Phys. Rev. A {\bf 51}, 1382
(1995); P. A. Ruprecht, M. J. Holland, K. Burnett, and M. Edwards,
{\it ibid.}   { \bf 51}, 4704 (1995); 
S. K. Adhikari, Phys. Rev. E {\bf 63}, 056704 (2001).

\bibitem{3b}B. I. Schneider and D. L. Feder,
{Phys. Rev. A}   { \bf 59}, 2232 (1999).


\bibitem{4} 
M. Edwards, P. A. Ruprecht, K. Burnett, R. J.
Dodd, and C. W. Clark, Phys. Rev. Lett. {\bf 77}, 1671 (1996);  M. J. Holland,
D. S. Jin, M. L. Chiofalo, and J. Cooper,
{\it ibid.}   {\bf 78},  3801 (1997).


\bibitem{11} F. Dalfovo, S. Giorgini, L. P. Pitaevskii, and S. Stringari,
Rev. Mod.  Phys. {\bf 71}, 463 (1999).  




 
 
\bibitem{2} Yu. Kagan, A. E. Muryshev, and G. V.  Shlyapnikov, { Phys.
Rev. Lett.} {\bf 81}, 933 (1998); M. Ueda and A. J.  
Leggett, { \it ibid.} {\bf 80}, 1576 (1998); 
V. S. Filho, A. Gammal, T.  Frederico,
and L. Tomio, { Phys.  Rev. A} {\bf 62}, 033605 (2000);  
M. Ueda and K.  
Huang,  { \it ibid.}  {\bf 61}, 043601 (1999); 
F. Kh.
Abdullaev, A. Gammal, L. Tomio, and T. Frederico,
   { \it ibid.}  {\bf 63}, 043604 (2001); 
R. J. Dodd, M.
Edwards, C. J. Williams, C. W. Clark, M. J. Holland, P. A. Ruprecht, and
K. Burnett,  { \it ibid.}  {\bf 54}, 661
 (1996); M. Houbiers and H. T. C. Stoof, {\it ibid.}   {\bf 54}, 5055
(1996); 
A. Eleftheriou and K. Huang,  {\it ibid.}   {\bf 61}, 043601
(2000); L. Berg\'e, T. J. Alexander, and Y. S. Kivshar, {\it ibid.}   {\bf
62},
023607 (2000);
S. K. Adhikari, { Phys.
Lett. A} {\bf 281}, 265 (2001); 
Phys. Rev. A {\bf
63}, 043611 (2001); J. Phys. B {\bf 34}, xxxx (2001); 
A. 
  Gammal, T. Frederico,  L. Tomio,  and F. Kh. Abdullaev, 
  { Phys. Lett. A} {\bf 267}, 305 (2000);
M.  Wadati  and T. Tsurumi,   {\it ibid.}    {\bf 247},  287
 (1998).
 
 

 
 



\bibitem{8}E. P. Gross, Nuovo Cimento {\bf 20}, 454 (1961);  L. P.
Pitaevskii, Zh. Eksp. Teor. Fiz. {\bf 40}, 646 (1961)[Sov.  Phys.  JETP
{\bf 13}, 451 (1961)]. 



\bibitem{2a}F. Dalfovo and S. Stringari,  Phys. Rev. A {\bf
53},
2477 (1996).

\bibitem{2b1}M. Holland and J. Cooper,  Phys. Rev. A {\bf
53},
R1954 (1996); F. Dalfovo and M. Modugno,  {\it ibid.}   {\bf
61},
023605 (2000).



\bibitem{2c}B. Jackson, J. F. McCann, and C. S. Adams,  Phys. Rev. A {\bf
61},
013604 (1999).



\bibitem{2d}
D. L. Feder, C. W. Clark, and B. I. Schneider,
Phys. Rev. Lett.  {\bf
82},
4956 (1999).

\bibitem{2e1} 
G. M. Kavoulakis, B. Mottelson, and C. J. Pethick, {Phys. Rev. A}
{\bf
62},
063605 (2000).


\bibitem{2e2}
J. J. Garc\'ia-Ripoll and
V. M. P\'erez-Garc\'ia; Phys. Rev. Lett.  {\bf
84},
4264 (2000);
V. M. P\'erez-Garc\'ia and J. J. Garc\'ia-Ripoll, Phys. Rev. A {\bf
62},
033601 (2000).


\bibitem{str}
F. Dalfovo and S. Stringari, Phys. Rev. A {\bf 63}, 011601
(2001); A. Recati, F. Zambelli, S. Stringari, Phys. Rev. Lett.  {\bf 86},
377 (2001);  
D. Guery-Odelin and S. Stringari, {\it ibid.} {\bf 83}, 4452 (1999);
S. Stringari, {\it ibid.}  {\bf 82}, 4371 (1999).     


\bibitem{w} J. E. Williams and M. J. Holland, Nature (London) {\bf 401},
568
(1999).
\bibitem{br}D. A. Butts and D. S. Rokshar,  Nature (London) {\bf 397}, 327
(1999).
\bibitem{exp1} K. W. Madison, F. Chevy, W. Wohlleben, and J. Dalibard,
Phys. Rev. Lett.  {\bf 84}, 806 (2000);  K. W. Madison, F. Chevy, V.
Bretin, and J. Dalibard, {\it ibid.} {\bf 86}, 4443 (2001).

\bibitem{exp2}  M. R.
Matthews, B. P. Anderson, P. C. Haljan, D. S. Hall, C. E. Wieman, E. A.
Cornell, {Phys. Rev. Lett.} {\bf 83}, 2498 (1999).

\bibitem{2f}R. J. Marshall, G. H. C. New, K. Burnett, and S. Choi, Phys. Rev. A
{\bf 59}, 2085 (1999).

\bibitem{dum} R. Dum, J. I. Cirac, M. Lewenstein, and P. Zoller,
Phys. Rev. Lett.  {\bf 80}, 2972 (1998).

\bibitem{2g} A. A. Svidzinsky and A. L. Fetter, Phys. Rev. A  {\bf
62},
063617 (2000).
\bibitem{exp3}
S. L. Cornish, N. R. Claussen, J. L. Roberts, E. A. Cornell, and
C. E. Wieman, Phys. Rev. Lett. 
{\bf 85}, 1795 (2000). 


\bibitem{hu}E. Timmermans, P. Tommasini, M. Hussein, and A. Kerman,
Phys. Rep. {\bf 315}, 199 (1999).

\bibitem{koo} S. E. Koonin and D. C. Meredith, Computational Physics Fortran Version,
Addison-Wesley Pub. Co., Reading, 1990, pp 169-180. 








\bibitem{xx1}A. Gammal, T. Frederico, and L. Tomio,  
 Phys. Rev. A {\bf 64}, in press (2001).
  






\bibitem{2b2} M. M. Cerimele, M. L. Chiofalo, F. Pistella, S. Succi, and
M. P. Tosi, Phys. Rev. E {\bf 62}, 1382 (2000). 



\bibitem{pw}L. Pauling and E. B. Wilson, {\it Introduction to Quantum
Mechanics,} McGraw-Hill, New York, 1935, pp 105-111.

\bibitem{hp}W. V. Houston and G. C. Phillips, {\it Principles of Quantum
Mechanics,} North-Holland, Amsterdam, 1973, pp 70-73. 



\bibitem{xxx}A. Gammal, T. Frederico, and L. Tomio, Phys. Rev. E {\bf 60},
2421 (1999).


\bibitem{yyy}S. K. Adhikari, Phys. Lett. A
{\bf 265} 91 (2000);
Phys. Rev. E   {\bf 62}, 2937 (2000).




\bibitem{yy1}J. L. Roberts, N. R. Claussen, S. L. Cornish, E. A. Donley,
E. A. Cornell,
C. E. Wieman, Phys. Rev. Lett. {\bf 86}, 4211 (20001).







\end{references}
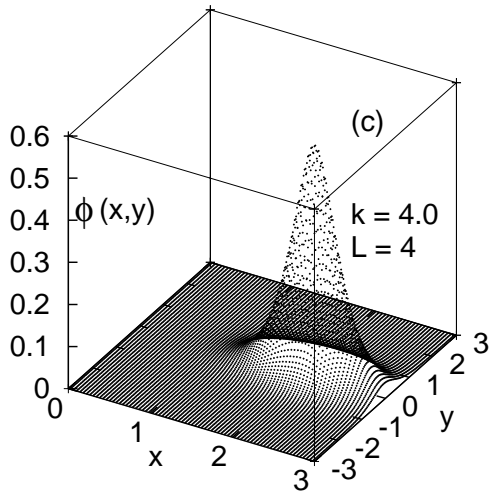
\end{document}